\documentclass[conference]{IEEEtran}
\IEEEoverridecommandlockouts
\usepackage{cite}
\usepackage{amsmath,amssymb,amsfonts}
\usepackage{algorithmic}
\usepackage{graphicx}
\usepackage{textcomp}
\usepackage{xcolor}
\usepackage{subscript}
\usepackage{lipsum}

\def\BibTeX{{\rm B\kern-.05em{\sc i\kern-.025em b}\kern-.08em
    T\kern-.1667em\lower.7ex\hbox{E}\kern-.125emX}}

\begin{document}

\title{Towards Secure and Interoperable Data~Spaces~for~6G: The 6G-DALI Approach\\
\thanks{This work has received funding from the Smart Networks and Services Joint Undertaking (SNS JU) under the European Union’s Horizon Europe Research and Innovation programme under Grant Agreement No 101192750.}
}

\author{
\IEEEauthorblockN{Dimitrios Amaxilatis,\\Themistoklis Sarantakos,\\Nikolaos Tsironis}
\IEEEauthorblockA{\textit{Spark Works Ltd.}\\
Galway, Ireland \\
\{d.amaxilatis,tsarantakos,ntsironis\}@sparkworks.net\\
0000-0001-9938-6211,0000-0002-7517-6997,\\0009-0009-8084-613X}
\and
\IEEEauthorblockN{Vasileios Theodorou}
\IEEEauthorblockA{\textit{Intracom Telecom}\\
Athens, Greece \\
theovas@intracom-telecom.com\\
}
\and
\IEEEauthorblockN{Christos Verikoukis}
\IEEEauthorblockA{\textit{Industrial Systems Institute,}\\
\textit{Athena Research
Center}\\
Patra, Greece \\
cveri@isi.gr\\
}
}

\maketitle

\begin{abstract}
The next generation of mobile networks, 6G, is expected to enable data-driven services at unprecedented scale and complexity, with stringent requirements for trust, interoperability, and automation.
Central to this vision is the ability to create, manage, and share high-quality datasets across distributed and heterogeneous environments. 
This paper presents the data architecture of the 6G-DALI project, which implements a federated dataspace and DataOps infrastructure to support secure, compliant, and scalable data sharing for AI-driven experimentation and service orchestration. 
Drawing from principles defined by GAIA-X and the International Data Spaces Association (IDSA), the architecture incorporates components such as federated identity management, policy-based data contracts, and automated data pipelines.
We detail how the 6G-DALI architecture aligns with and extends GAIA-X and IDSA reference models to meet the unique demands of 6G networks, including low-latency edge processing, dynamic trust management, and cross-domain federation.
A comparative analysis highlights both convergence points and necessary innovations.

\end{abstract}


\section{Introduction}


The advent of sixth-generation (6G) wireless networks promises to revolutionize telecommunications by enabling unprecedented data rates, ultra-low latency communication, and massive device connectivity that will support transformative applications, including extended reality, autonomous systems, and intelligent industrial automation. Unlike previous network generations that primarily focused on connectivity improvements, 6G represents a paradigm shift toward data-centric network architectures where the value lies not merely in transmitting information, but in intelligently processing, sharing, and monetizing the vast volumes of heterogeneous data generated across the network ecosystem. This fundamental transformation necessitates novel approaches to data management that can handle the exponential growth in data volume and velocity while ensuring security, sovereignty, and interoperability across diverse stakeholders.

The complexity of 6G data management stems from the unprecedented diversity of data sources, ranging from terrestrial and satellite networks to edge devices and artificial intelligence applications, each with unique characteristics, quality requirements, and processing demands. Traditional data management approaches prove inadequate for addressing the real-time analytics requirements, federated learning capabilities, and cross-domain collaboration essential for 6G network optimization and service delivery. Furthermore, the multi-stakeholder nature of 6G development, involving telecommunications operators, equipment vendors, research institutions, and vertical industry players, introduces complex requirements for secure data sharing while preserving data sovereignty, intellectual property rights, and regulatory compliance across multiple jurisdictions.

European initiatives such as GAIA-X~\cite{aisbl_gaia-x_2022} and the International Data Spaces Association (IDSA)~\cite{international_data_spaces_association_ids_2022} have established foundational frameworks for secure and sovereign data sharing, emphasizing principles of trust, interoperability, and participant control over data assets. These frameworks provide essential governance mechanisms and architectural patterns that can be adapted for domain-specific applications while maintaining compatibility with broader data ecosystem initiatives. However, the specific requirements of 6G telecommunications networks—including massive scale, real-time processing, heterogeneous integration, and specialized AI/ML workloads—necessitate domain-optimized implementations that can leverage established frameworks while addressing challenges of next-generation network environments.

This paper presents the 6G-DALI approach, which develops a comprehensive data space architecture specifically designed for 6G telecommunications research and experimentation. Our approach combines the proven governance and trust frameworks of GAIA-X and IDSA with domain-specific innovations tailored to the unique requirements of 6G network environments, including specialized components for radio access network (RAN) model management, testbed integration, and real-time data processing at scale. 
The 6G-DALI architecture demonstrates how European data space principles can be successfully adapted for telecommunications while maintaining interoperability with broader data ecosystem initiatives, providing a foundation for secure, sovereign, and collaborative 6G network development that enables stakeholders to share data and AI/ML models while preserving control over their intellectual property and ensuring compliance with regulatory requirements.

\section{Related Work}




The development of data spaces for telecommunications networks has gained significant attention as 6G research intensifies, with various initiatives exploring different approaches to address the unique challenges of next-generation network data management.
European data space initiatives have established foundational frameworks that influence telecommunications-specific implementations, while emerging research demonstrates the growing recognition of data-centric architectures across multiple domains.

The GAIA-X initiative represents the most comprehensive European effort to establish federated data infrastructures based on sovereignty and trust principles. 
GAIA-X aims to create a federated open data infrastructure based on European values regarding data and cloud sovereignty, implementing a data sharing architecture with common standards, best practices, tools, and governance mechanisms.
The GAIA-X Trust Framework operationalizes requirements through verifiable credentials and linked data representations, enabling machine-readable trusted information exchange.
The initiative has successfully demonstrated application across various sectors including healthcare, manufacturing, and transportation, establishing patterns for secure data sharing while maintaining participant sovereignty over their data assets.

The IDSA has developed the most mature reference architecture for secure data sharing through its Reference Architecture Model (IDS-RAM). The IDS-RAM provides a framework for technically enforced agreements for data sharing, enabling data exchange within trusted ecosystems while preserving data sovereignty through its technology-agnostic and domain-agnostic information model. The five-layer IDS-RAM structure has influenced numerous implementations across industries, with successful deployments in automotive, manufacturing, and logistics sectors demonstrating the practical applicability of its architectural principles for complex multi-stakeholder environments requiring secure data sharing mechanisms.

The Data Spaces Support Centre (DSSC)~\cite{dssc} has complemented these foundational initiatives by providing practical guidance and blueprints for data space implementation. The Data Spaces Blueprint v2.0~\cite{dssc_blueprint_2024} presents a comprehensively updated reference architecture for building sovereign data spaces, with multi-layered architecture models that clearly separate logical, functional, and technical levels. Various sectoral data spaces have emerged following these blueprints, including the European Health Data Space (EHDS), Manufacturing Data Space, and Mobility Data Space, each adapting the core principles to domain-specific requirements while maintaining interoperability with the broader European data ecosystem.

Domain-specific data space implementations have demonstrated the adaptability of European frameworks across diverse sectors. The Manufacturing Data Space initiative has shown how industrial data can be shared securely across supply chains while preserving competitive advantages. The European Open Science Cloud (EOSC) has implemented data space principles for scientific research collaboration, enabling researchers to share datasets and computational resources across institutional boundaries. These implementations have established proven patterns for data sovereignty, federated identity management, and policy-based access control that can be adapted for telecommunications applications.

The emergence of Common European Data Spaces (CEDS)~\cite{CEDS2020} has created momentum for operational data space solutions across multiple domains. The European Industrial Data Space (IDS)~\cite{EuropeanIndustrialDataSpace} has demonstrated large-scale implementation of federated data sharing in manufacturing environments, while sectoral initiatives in agriculture, energy, and finance have shown the versatility of data space architectures for handling diverse data types and regulatory requirements. These initiatives collectively demonstrate the maturity of European data space frameworks and their readiness for application in specialized domains such as telecommunications.

Technical infrastructure for data spaces has evolved significantly, with the Eclipse Data Space Connector (EDC)~\cite{eclipse_edc_2024} providing open-source implementations of IDSA specifications and GAIA-X compliance frameworks. Multiple cloud vendors have integrated data space capabilities into their platforms, enabling organizations to participate in federated data ecosystems while maintaining control over their infrastructure choices. The development of standardized APIs and protocols has facilitated interoperability between different data space implementations, reducing barriers to cross-domain collaboration.

Regulatory frameworks have increasingly emphasized the importance of data spaces for European digital sovereignty. The European Commission's Data Act~\cite{EUDataAct2023} and Digital Services Act~\cite{DigitalServicesAct2022} provide legal foundations that data space initiatives must address, emphasizing data portability, interoperability, and fair access principles. The Data Governance Act~\cite{DataGovernanceAct2022} establishes frameworks for data intermediation services that align closely with data space architectural principles, providing regulatory clarity for cross-border data sharing initiatives.

Despite the success of data spaces across various sectors, telecommunications presents unique challenges that existing implementations have not fully addressed. The real-time processing requirements, massive scale, and complex multi-stakeholder environments characteristic of telecommunications networks require specialized adaptations of data space principles. The need for AI/ML model sharing, experimental data generation, and testbed federation introduces requirements that go beyond traditional data sharing scenarios, highlighting opportunities for telecommunications-specific innovations within the broader European data space ecosystem.
%

\section{The 6G-DALI Dataspace}

\subsection{6G-DALI}


The 6G-DALI project aims to deliver a revolutionary end-to-end AI framework for 6G networks, structured around two interdependent pillars: AI experimentation as a service via MLOps and data analytics collection and storage via DataOps. The project's primary objective is to address the critical gap in AI/ML adoption for 6G systems by providing a user-friendly framework that connects 6G data with verticals, ML developers, and experimenters. By leveraging established 6G testbeds from the flagship SNS C phase 2 (SUNRISE-6G~\cite{SUNRISE6G}) project, 6G-DALI seeks to democratize AI experimentation in telecommunications, enabling automated ML model lifecycle management, including training, validation, finetuning, hyperparameter optimization, and performance benchmarking across heterogeneous 6G infrastructures.


The 6G-DALI data lifecycle is orchestrated through an innovative Extract, Load, Transform (ELT) pipeline that distinguishes between "cold data" (existing datasets in the 6G Dataspace) and "hot data" (generated on-demand through testbed experiments). The framework employs intent-driven DataOps, where users can formulate data requests using natural language that are automatically translated by Large Language Models (LLMs) into executable experiments or dataset queries. When datasets are unavailable in the dataspace, the system intelligently selects optimal testbeds for experiment execution, applies AI-driven data cleaning and augmentation techniques during the transformation phase, and ensures high-quality datasets are made available for ML model training. This comprehensive approach addresses the critical challenge of data scarcity in 6G research while maintaining data quality and accessibility standards.


Central to the 6G-DALI architecture is the establishment of a 6G Dataspace built following IDS and Gaia-X guidelines, ensuring secure, sovereign, and interoperable data sharing across European testbeds and cloud vendors. This dataspace serves as the foundation for collaborative AI workflows, enabling seamless integration of multiple 6G testbeds through standardized adapters and connectors. The dataspace not only facilitates data discovery and sharing but also supports automated experimentation workflows where ML developers can access pre-trained models, high-quality datasets, and staging environments for model validation. By implementing trustworthy AI mechanisms including drift detection, explainable AI techniques, and uncertainty quantification, the dataspace ensures reliable and transparent AI model deployment across the entire 6G ecosystem.


The 6G-DALI framework represents a paradigm shift toward Native AI integration in telecommunications, offering groundbreaking innovations including collaborative Transfer Learning, automated Reinforcement Learning Operations (RLOps), and Federated Learning Operations (FLOps) leveraging the 3GPP Network Data Analytics Function (NWDAF). The project's Digital Twin Testbed enables large-scale experimentation scenarios that traditional testbeds cannot support, while the meta-orchestration solution provides hardware-agnostic deployment capabilities across the cloud-edge continuum. Through its comprehensive approach to AI experimentation automation, 6G-DALI not only accelerates 6G research and development but also establishes the foundation for sustainable AI-driven network management and optimization, ultimately contributing to Europe's technological sovereignty in next-generation telecommunications.

\subsection{Service Overview}

\begin{figure}[h]
    \centering
    \includegraphics[width=0.9\linewidth]{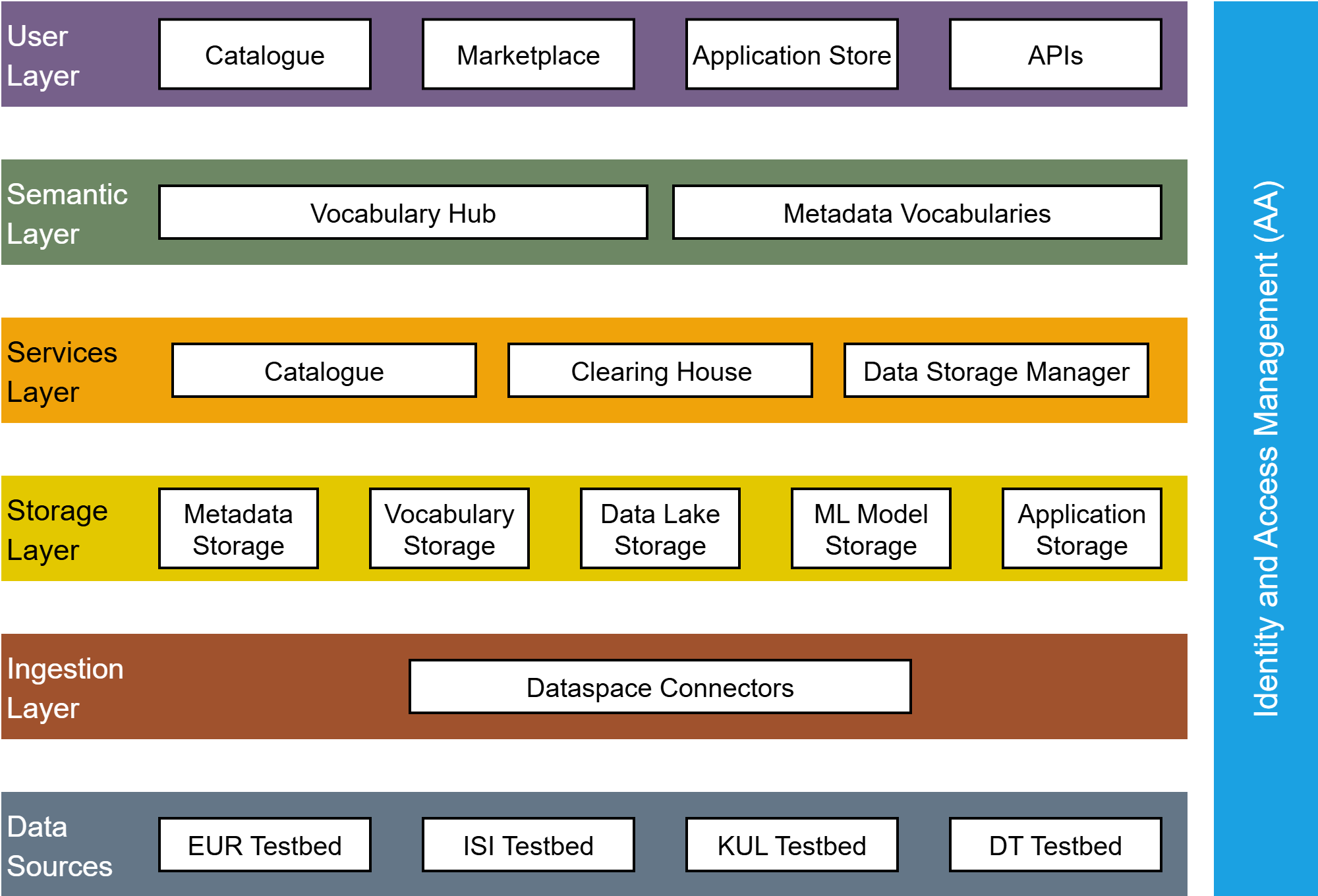}
    \caption{Service layers of the 6G-DALI Data Space and Data Lake.}
    \label{fig:services}
\end{figure}

Figure~\ref{fig:services} depicts the six service layers of the 6G-DALI Data Space. They are designed to facilitate trusted, interoperable, and secure data sharing across the telecommunications ecosystem, supporting the dynamic needs of 6G networks, integrating data, AI/ML models, and applications within a federated, standards-compliant data space.

At the base of the stack lies the \textit{Data Sources layer}, comprising diverse 6G testbeds that act as primary data providers for the ecosystem. These include EUR Testbed, ISI Testbed, KUL Testbed, and DT Testbed, which are provided by project partners for now, but more organizations will be invited to join as the system matures.
Above this, the \textit{Ingestion layer} hosts the Dataspace Connectors, enabling integration of heterogeneous data sources. These connectors perform ingestion, transformation, and protocol mediation, ensuring data from each testbed can be shared securely and in compliance with GAIA-X and IDSA principles to the common 6G Data Space.
The \textit{Storage layer} provides a distributed and modular foundation for managing diverse data assets and models along with their metadata. It includes:

\begin{itemize}
\item Metadata Storage for dataset descriptions,
\item Vocabulary Storage for semantic vocabularies and ontologies that underpin interoperability,
\item Data Lake Storage as the main scalable repository for processed data
\item ML Model Storage for machine learning models and their artifacts,
\item Application Storage for app-specific configurations and artifacts.
\end{itemize}

Sitting above the storage layer, the \textit{Services layer} delivers key federation-wide services. 
The Catalogue enables discovery, registration, and management of data assets and ML models across the 6G data space. The Clearing House provides essential governance functions: access control, transaction management, usage tracking, and compliance auditing, supporting secure, transparent data exchange. The Data Storage Manager coordinates and optimizes data placement and retrieval across the distributed storage infrastructure.

The \textit{Semantic layer} ensures true interoperability and harmonization across data, models, and applications. It includes the Vocabulary Hub for managing and distributing semantic resources and Metadata Vocabularies for establishing standardized data descriptions that enable seamless integration and automated reasoning.

At the top, the \textit{User layer} provides multiple interaction channels for accessing the system: a Catalogue for data and ML model discovery, a Marketplace for trusted data trading, monetization, and licensing, an Application Store for accessing AI/ML-powered services and data transformation applications, as well as APIs for programmatic interaction by developers and external systems.

Finally, Identity and Access Management (IAM) functions span all layers, ensuring authentication, authorization, and fine-grained access control. This guarantees secure and privacy-preserving data sharing while enabling collaborative experimentation, AI/ML model development, and advanced service delivery critical for 6G.

\subsection{Architecture}




The architecture of the \textit{6G-DALI Data Space} represents a federated, secure, and semantically rich framework designed to support trusted data sharing and AI-driven experimentation in 6G networks.
At the core of this architecture is the interplay between the \textit{Data Space}, where data assets, services, and models are described, governed, and shared, and the \textit{Data Lake}, where actual datasets and models are ingested, stored, and processed at scale.

The \textit{Data Space} contains a \textit{catalogue} that registers datasets, services, machine learning models, and radio access network (RAN) models.
It is integrated with a \textit{catalogue UI} \& a \textit{marketplace}, enabling discovery of the available assets.
An \textit{App Store} further extends the functionality by providing access to applications and tools within the Data Space ecosystem.
Critical to ensuring trust and compliance are the \textit{clearing house} and \textit{governance} modules, which support transaction logging, auditability, and enforcement of data usage policies.
Semantic interoperability within the Data Space is achieved through the \textit{vocabulary hub} and schema layer.
Vocabularies and schemas define the semantics of shared data and services, supporting consistent metadata descriptions and alignment across domains.
These definitions are stored in a dedicated semantic database (triplestore) that underpins the storage layer of the Data Space.

On the Data Lake side, ingestion is managed via \textit{dataspace connectors}, which apply policy controls during data transfer.
The \textit{Data Lake backend} is responsible for handling the storage and management of raw data, machine learning models, and associated metadata.
Data and models are organized according to schemas defined, with information stored in the internal database and object storage systems providing scalability and compatibility with cloud-native storage standards.

Authentication and authorization are central services in this architecture, providing federated identity management and access control mechanisms that ensure only authorized actors can interact with the \textit{Data Space} and \textit{Data Lake} components.
The \textit{dataspace connectors} act as the trusted enforcement points for these controls, facilitating secure, policy-driven data exchange between internal services, external testbeds, and other federated participants.

All these components are designed to provide the 6G-Dali Cloud Services user-facing components with all the information needed to manage and discover datasets and ML models needed for the execution of 6G-related experiments.
These environments connect securely to the Data Space and Data Lake via their own \textit{dataspace connectors}, ensuring that all interactions with data comply with governance policies and contractual obligations.

In summary, the 6G-DALI Data Space architecture enables interoperable, secure, and governed data sharing across distributed domains, with strong alignment to the principles of \textit{GAIA-X} and \textit{IDSA}.
It combines rich metadata and governance capabilities in the Data Space with scalable, flexible data storage and processing in the Data Lake, supporting the advanced data needs of 6G experimentation and AI/ML workflows.

\begin{figure}[h]
    \centering
    \includegraphics[width=0.95\linewidth]{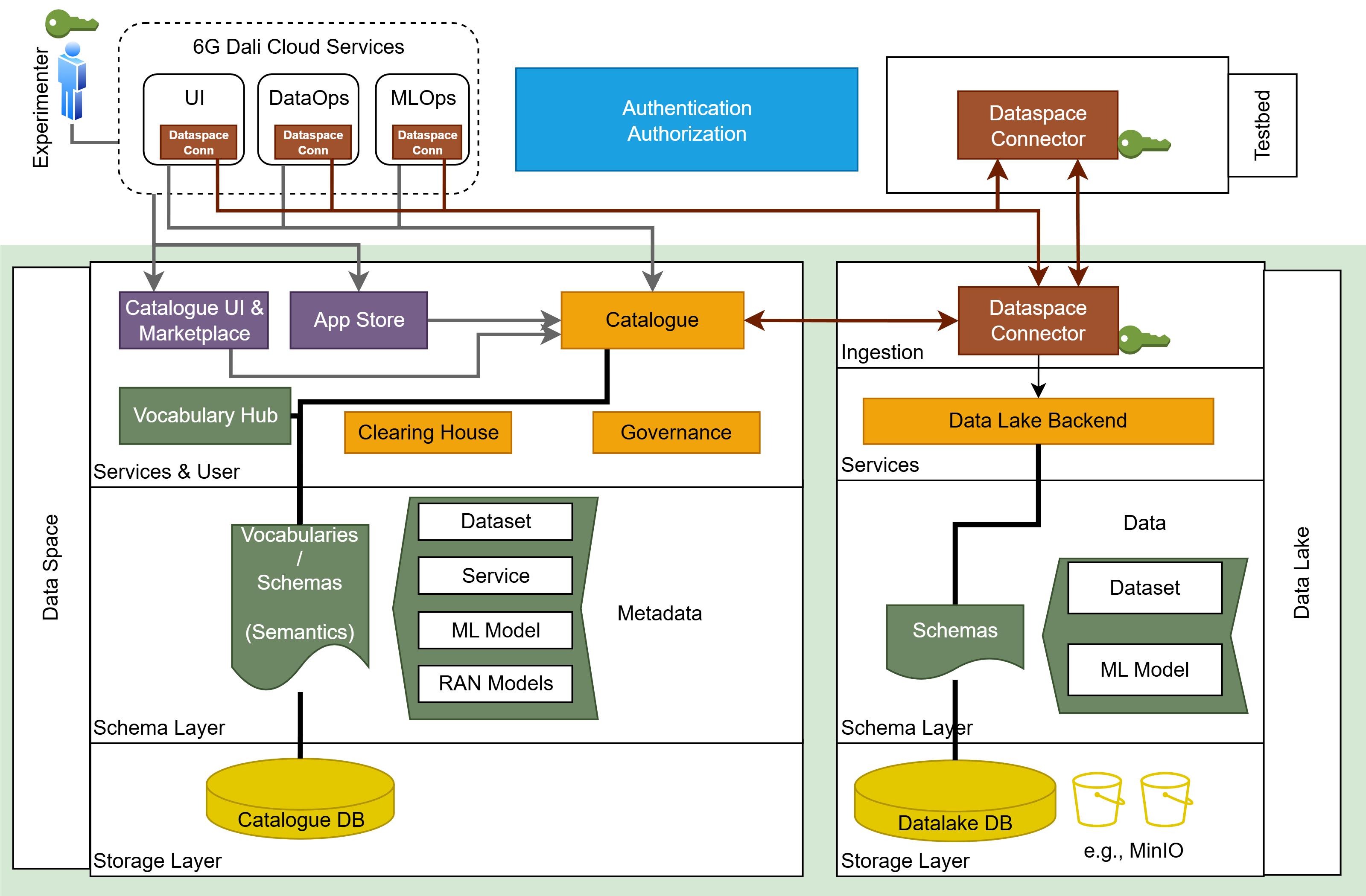}
    \caption{Architecture of the 6G-DALI Data Space and Data Lake.}
    \label{fig:architecture}
\end{figure}

\section{Comparison with GAIA-X and IDSA Reference Architectures}



\subsection{Gaia-X}
Our architecture is strongly aligned with GAIA-X principles while introducing domain-specific adaptations for 6G telecommunications environments. At its core, both architectures prioritize federated trust and identity management based on Self-Sovereign Identity (SSI) principles and W3C Verifiable Credentials, ensuring that interoperability is guaranteed while preserving participant sovereignty over their digital identities.

We have designed our data sovereignty and usage control mechanisms to align with GAIA-X's approach, where participants maintain full self-determination over their data exchange and sharing. Our emphasis on policy-driven data exchange through dataspace connectors and clearinghouse mechanisms for transaction logging and compliance auditing directly corresponds to GAIA-X's data sovereignty services. Both systems implement sophisticated governance frameworks that ensure data usage policies are enforced and auditable, though we have tailored these mechanisms to the experimental nature of 6G telecommunications research.

Our catalogue-based discovery mechanisms reflect the same commitment to semantic interoperability found in GAIA-X. While GAIA-X's Federated Catalogue constitutes an indexed repository of Self-Descriptions to enable discovery and selection of providers and their service offerings, we have implemented a catalogue system that registers datasets, services, ML models, and RAN models, integrated with marketplace functionality. This similarity extends to semantic frameworks, where GAIA-X emphasizes controlled vocabularies and SKOS-compatible ~\cite{noauthor_skos_nodate} semantic frameworks, directly corresponding to our semantic layer with its Vocabulary Hub and Metadata Vocabularies for standardized data descriptions.

However, we have introduced several domain-specific innovations that distinguish our approach from the broader GAIA-X framework. While GAIA-X targets a broad European data ecosystem federation across multiple industries, we have specifically architected our system for telecommunications and 6G network experimentation, incorporating specialized components like RAN model management and testbed integration. Our explicit separation of the Data Space (metadata, governance, discovery) from the Data Lake (actual data storage and processing) represents a significant architectural decision that differs from GAIA-X's unified data space concepts, enabling more efficient handling of the large-scale data processing requirements typical in telecommunications research.

Our storage layer specialization reflects these domain-specific requirements, including dedicated ML Model Storage and testbed-specific data management using technologies like MinIO, while GAIA-X provides more generic resource abstractions for physical and virtual resources. Additionally, we have explicitly incorporated multiple testbeds as data sources, reflecting the experimental nature of 6G research, whereas GAIA-X focuses on broader participant onboarding processes without this testbed-specific integration.

Our marketplace integration also represents a notable architectural choice, embedding marketplace functionality directly into the user layer alongside the catalogue, while GAIA-X positions marketplaces as domain-specific extensions rather than core architectural components. This design decision reflects our emphasis on facilitating data monetization and trading within the telecommunications research community.

Despite these differences, we have ensured that our architecture explicitly aligns with GAIA-X, IDSA, and EDIB~\cite{EDIB2024} frameworks, guaranteeing interoperability with broader European data space initiatives while maintaining domain-specific functionality. Both systems implement automated compliance processes and versioned rules, and both support open-source implementations and federated service composition, though we emphasize AI/ML model deployment and telecommunications-specific services. This approach allows us to successfully adapt GAIA-X's foundational principles of trust, sovereignty, and interoperability to the specific requirements of 6G telecommunications research, creating a domain-optimized implementation that maintains compliance with European data space standards while addressing the unique challenges of network experimentation and AI/ML model sharing in telecommunications environments.

\subsection{IDSA RAM}

Alignment with the IDS-RAM principles is also strong while introducing domain-specific optimizations for 6G telecommunications environments. 
Both frameworks share a fundamental commitment to data sovereignty, where data sharing occurs peer-to-peer with data owners maintaining control and responsibility for their data through policy enforcement. 

Our implementation of \textit{dataspace connectors} directly corresponds to IDS Connectors that initiate data exchange from and to internal data resources and enterprise systems of participating organizations, ensuring that data sovereignty is maintained throughout the exchange process.

The layered architectural approach represents a core convergence between our system and IDS-RAM.
IDS-RAM employs a five-layer structure (business, functional, process, information, and system) to express various stakeholders' concerns and viewpoints at different levels of granularity, while our architecture implements a six-layer approach that maps closely to these concepts.
Our Data Sources layer aligns with its business layer participant roles, our Services layer corresponds to the functional layer requirements, and our storage layer maps to the system layer technical architecture. 
The ingestion layer with dataspace connectors directly implements IDS-RAM's connector-based approach for secure data exchange within trusted ecosystems while preserving data sovereignty.

We have designed comprehensive governance and trust mechanisms that align with IDS-RAM principles.
IDS-RAM addresses security, privacy, trust, and governance through guidelines for authentication, authorization, data protection, and compliance with regulatory requirements, which aligns with our Identity and Access Management (IAM) spanning all layers and our clearing house mechanisms for transaction logging and compliance auditing. Our vocabulary hub and semantic layer implementation reflects IDS-RAM's information model that enables comprehensive description of data assets and interoperability required for trusted data exchange

Our catalogue and discovery mechanisms demonstrate shared architectural principles with IDS-RAM. 
While IDS-RAM's Metadata Broker stores information about data endpoints offered by participants and provides query interfaces for connectors, we have implemented a catalogue system that registers datasets, services, ML models, and RAN models with integrated marketplace functionality. 
Sophisticated metadata management is also implemented to enable the discovery and selection of appropriate data assets and services.

However, our architecture introduces several telecommunications-specific innovations that distinguish it from the generic IDS-RAM framework. 
While IDS-RAM is designed for cross-sector data exchange with standardized roles and interactions, we have specifically architected our system for 6G network experimentation, incorporating specialized components like RAN model management and explicit testbed integration.
Our dual Data Space and Data Lake architecture represents a significant adaptation from IDS-RAM's connector-centric approach, enabling more efficient handling of the massive data processing requirements typical in telecommunications research.
The storage specialization in our architecture reflects these domain-specific requirements. 
We implement dedicated ML Model Storage and testbed-specific data management using technologies like MinIO, while IDS-RAM provides generic system layer specifications for technical architecture and integration points. 
Our explicit separation of metadata governance (Data Space) from actual data storage and processing (Data Lake) enables optimized performance for the high-volume, high-velocity data characteristics of 6G experimentation.

Our marketplace integration approach also differs from IDS-RAM's model. 
We have embedded marketplace functionality directly into the user layer alongside the catalogue, facilitating immediate data monetization and trading within the telecommunications research community. 
This contrasts with IDS-RAM's approach, where metadata brokers provide query interfaces but marketplace functionality is typically implemented as separate components.
Despite these domain-specific adaptations, we maintain full compatibility with IDS-RAM's trust, certification, and governance perspectives and ensure our implementation supports the standardized Dataspace Protocol developed by IDSA for ensuring interoperability and trust across data spaces. 
Our architecture demonstrates how IDS-RAM's technology-agnostic and domain-agnostic information model can be successfully specialized for telecommunications while maintaining interoperability with broader data space ecosystems. 
This approach allows us to leverage the mature governance and trust frameworks of IDS-RAM while addressing the unique challenges of 6G network experimentation, AI/ML model sharing, and real-time data processing requirements in telecommunications research environments.

\section{Conclusion}
In this paper, we presented the 6G-DALI Data Space architecture, which adapts European data space frameworks to the specific requirements of 6G telecommunications. Our approach extends GAIA-X and IDSA principles to domain-specific applications while ensuring interoperability with broader data ecosystems.

The core innovation lies in the dual Data Space and Data Lake design, separating metadata governance from large-scale data processing to meet the demands of 6G research. By integrating RAN model management, testbed connectors, and policy-based governance, 6G-DALI supports secure, interoperable, and AI-driven experimentation across federated environments.

Beyond research, the architecture offers practical benefits for industry and academia by enabling sovereign, cross-domain data sharing and reducing the cost of AI model development. Its alignment with GAIA-X and IDSA ensures compatibility with emerging European data spaces and facilitates adoption within real-world 6G initiatives.

Future work will extend validation across additional testbeds and cloud providers, integrate GAIA-X compliant credentialing, and benchmark DataOps and MLOps workflows at scale. These efforts will establish concrete performance indicators for scalability, interoperability, and trust, paving the way for a fully operational 6G data and AI experimentation ecosystem.

\bibliographystyle{IEEEtran}
\bibliography{conference_101719}

@misc{aisbl_gaia-x_2022,
	title = {Gaia-{X} {Architecture} {Document} – {Version} 3.0},
	url = {https://gaia-x.eu/resources/gaia-x-architecture-document},
	author = {AISBL, Gaia-X.},
	year = {2022},
}

@misc{international_data_spaces_association_ids_2022,
	title = {{IDS} {Reference} {Architecture} {Model} 4.0},
	url = {https://internationaldataspaces.org/publications/},
	author = {{International Data Spaces Association}},
	year = {2022},
}

@misc{dssc,
  author = {{Data Spaces Support Centre}},
  title = {Data Spaces Documentation and Guidelines},
  year = {2024},
  url = {https://dssc.eu/},
  note = {Accessed: 21/07/2025},
}

@misc{dssc_blueprint_2024,
  author = {{Data Spaces Support  Centre}},
  title = {Data Spaces Blueprint v2.0: A Reference Architecture for Sovereign Data Spaces},
  year = {2024},
  url = {https://dssc.eu/blueprint/},
  note = {Accessed: 21/07/2025},
}

@misc{eclipse_edc_2024,
  author = {{Eclipse Foundation}},
  title = {Eclipse Dataspace Connector (EDC)},
  year = {2024},
  url = {https://github.com/eclipse-edc/Connector},
  note = {Accessed: 21/07/2025},
}

@misc{noauthor_skos_nodate,
	title = {{SKOS} {Simple} {Knowledge} {Organization} {System} - home page},
	url = {https://www.w3.org/2004/02/skos/},
    note = {Accessed: 21/07/2025},
}

@misc{EDIB2024,
  author = {{European Data Innovation Board}},
  title = {European Data Innovation Board},
  year = {2024},
  organization = {European Commission},
  url = {https://digital-strategy.ec.europa.eu/en/policies/data-innovation-board},
  note = {Accessed: 21/07/2025},
}

@misc{SUNRISE6G,
  author = {{SUNRISE-6G Consortium}},
  title = {SUNRISE-6G: Smart Networks in 6G for Universal, Reliable and Intelligent Service Enablement},
  year = {2022},
  organization = {European Union's Horizon Europe Research and Innovation Programme},
  note = {Grant Agreement No. 101096124},
  url = {https://sunrise-6g.eu/},
  note = {Accessed: 21/07/2025},
}

@misc{EUDataAct2023,
  author = {{European Parliament and Council}},
  title = {Regulation (EU) 2023/2854 on harmonised rules on fair access to and use of data and amending Regulation (EU) 2017/2394 and Directive (EU) 2020/1828 (Data Act)},
  year = {2023},
  month = {December},
  day = {13},
  journal = {Official Journal of the European Union},
  volume = {L 2023/2854},
  publisher = {Publications Office of the European Union},
  address = {Luxembourg},
  url = {https://eur-lex.europa.eu/eli/reg/2023/2854/oj}
}

@misc{DigitalServicesAct2022,
  author = {{European Parliament and Council}},
  title = {Regulation (EU) 2022/2065 of the European Parliament and of the Council on a Single Market For Digital Services and amending Directive 2000/31/EC (Digital Services Act)},
  year = {2022},
  month = {October},
  day = {19},
  journal = {Official Journal of the European Union},
  volume = {L 277/1},
  publisher = {Publications Office of the European Union},
  address = {Luxembourg},
  url = {https://eur-lex.europa.eu/eli/reg/2022/2065/oj}
}

@misc{DataGovernanceAct2022,
  author = {{European Parliament and Council}},
  title = {Regulation (EU) 2022/868 of the European Parliament and of the Council on European data governance and amending Regulation (EU) 2018/1724 (Data Governance Act)},
  year = {2022},
  month = {May},
  day = {30},
  journal = {Official Journal of the European Union},
  volume = {L 152/1},
  publisher = {Publications Office of the European Union},
  address = {Luxembourg},
  url = {https://eur-lex.europa.eu/eli/reg/2022/868/oj}
}

@misc{CEDS2020,
  author = {{European Commission}},
  title = {A European strategy for data},
  year = {2020},
  month = {February},
  day = {19},
  publisher = {European Commission},
  address = {Brussels},
  url = {https://eur-lex.europa.eu/legal-content/EN/TXT/?uri=CELEX\%3A52020DC0066},
  note = {Communication from the Commission to the European Parliament, the Council, the European Economic and Social Committee and the Committee of the Regions}
}

@misc{EuropeanIndustrialDataSpace,
  author = {{European Commission}},
  title = {European Industrial Data Space},
  year = {2022},
  organization = {European Commission, Directorate-General for Internal Market, Industry, Entrepreneurship and SMEs},
  url = {https://digital-strategy.ec.europa.eu/en/policies/data-spaces-manufacturing},
  note = {Common European Data Space for Manufacturing}
}

\end{document}